\documentclass[aps,prl,twocolumn,showpacs,superscriptaddress]{revtex4}

\def\bavs{BaVS$_{3}$}
\def\d1{$\textit{d}$$^{1}$}

\usepackage{graphicx}

\begin{document}

\title{Experimental Electronic Structure and Interband Nesting in \bavs}

\author{S. Mitrovic}\email{slobodan.mitrovic@epfl.ch}
\affiliation{Institut de Physique des Nanostructures, Ecole Polytechnique F\'ed\'erale de Lausanne, CH-1015 Lausanne, Switzerland}
\author{P. Fazekas}
\affiliation{Institut de Physique de la Mati\`ere Complexe, Ecole Polytechnique F\'ed\'erale de Lausanne, CH-1015 Lausanne, Switzerland}
\affiliation{Research Institute for Solid State Physics and Optics, Budapest, H-1525 Hungary}
\author{C. S$\o$ndergaard}
\affiliation{Institut de Physique des Nanostructures, Ecole Polytechnique F\'ed\'erale de Lausanne, CH-1015 Lausanne, Switzerland}
\author{D. Ariosa}
\author{N. Bari\v{s}i\'{c}}
\author{H. Berger}
\author{D. Clo\"{e}tta}
\author{L. Forr\'{o}}
\affiliation{Institut de Physique de la Mati\`ere Complexe, Ecole Polytechnique F\'ed\'erale de Lausanne, CH-1015 Lausanne, Switzerland}
\author{H. H\"ochst}
\affiliation{Synchrotron Radiation Center, University of Wisconsin - Madison, Stoughton,
Wisconsin Ê53589}
\author{I. Kup\v{c}i\'{c}}
\affiliation{Department of Physics, Faculty of Science, University of Zagreb, HR-10001 Zagreb, Croatia} \affiliation{Institut de
Physique de la Mati\`ere Complexe, Ecole Polytechnique
F\'ed\'erale de Lausanne, CH-1015 Lausanne, Switzerland}
\author{D. Pavuna}
\affiliation{Institut de Physique de la Mati\`ere Complexe, Ecole Polytechnique F\'ed\'erale de Lausanne, CH-1015 Lausanne, Switzerland}
\author{G. Margaritondo}
\affiliation{Institut de Physique des Nanostructures, Ecole Polytechnique F\'ed\'erale de Lausanne, CH-1015 Lausanne, Switzerland}
\affiliation{Institut de Physique de la Mati\`ere Complexe, Ecole Polytechnique F\'ed\'erale de Lausanne, CH-1015 Lausanne, Switzerland}

\date{\today}

\begin{abstract}

The correlated $3d$ sulphide \bavs{} is a most interesting compound because of the apparent coexistence of one-dimensional and three-dimensional properties. Our experiments explain this puzzle and shed new light on its electronic structure. High-resolution angle-resolved photoemission measurements in a 4eV wide range below the Fermi level explored the coexistence of weakly correlated $a_{1g}$ wide-band and strongly correlated
$e_g$ narrow-band $d$-electrons that is responsible for the complicated behavior of this material. The most relevant result is the evidence for $a_{1g}$--$e_g$ inter-band nesting
condition.

\end{abstract}



\maketitle

Fundamental arguments as well as the general interest in
functional nano-systems justify a broad and increasing interest in electronic phenomena
in one dimension. In this context, much attention has been devoted to the correlated
transition metal sulphide \bavs{} that offers the
puzzling combination of structural quasi-one-dimensionality with
three-dimensional (3D) character of some fundamental electrical and magnetic
properties. Reconciling these features is a challenge for
condensed matter physics.

The crystal structure can be envisaged as a triangular lattice of
chains of face-sharing VS$_{6}$ octahedra. Quasi-one-dimensionality
is apparent from the fact that V--V distances in the $a$--$b$
plane (6.72$\textrm{\AA}$) are almost 2.4 times as large as in the
$c$-direction of chains (2.84$\textrm{\AA}$). Naively, one would
expect good metallic conduction along the chains, and poor
conduction in the perpendicular directions. Indeed, this was the
prevalent view of the nature of \bavs{} as long as only
polycrystalline samples were available \cite{massenet}. The first
resistivity measurements on single crystal samples \cite{Mihaly00}
showed that in a wide range of temperatures, conductivity is
essentially isotropic ($\sigma_{c}/\sigma_{a} \sim 3$). This required a
re-examination of the effective dimensionality of \bavs{},
and its relationship to the correlation phenomena displayed by the
system.

A salient feature of \bavs{} is the second-order metal--insulator
transition (MIT) at T$_{\rm{MI}}$=69K, which is manifested also in
a strong susceptibility cusp. In most two- or three-dimensional
transition metal compounds, the MIT is
accompanied by a magnetic, or a structural transition. However,
\bavs{} is exceptional. The phase on the low-$T$ side of the MIT is
non-magnetic, and the structural aspect of the MIT was long
overlooked until the recent discovery of tetramerization
(doubling of the unit cell in the $c$-direction) \cite{inami}.

\bavs{} is a $3d^{1}$ system with intermediate strength of
correlations \cite{pb2002}. Local correlations are not strong
enough to sustain a paramagnetic insulating phase in high-quality
undoped samples \cite{Mihaly00,gauzzi}. The $T > T_{\rm{MI}}$ phase
is metallic but certainly a correlated metal: the susceptibility
is Curie-like, showing that a substantial fraction of the V sites
carries a local moment \cite{Graf95,Mihaly00}. The MIT should be
characterized as a Mott--Slater transition \cite{pb2002}: a
symmetry breaking transition assisted by correlations.
The fact that the MIT must be accompanied by symmetry breaking,
was deduced from the observation that the MIT remains a continuous
phase transition in a wide range of pressures
\cite{forro00,pb2002}, prior to the experimental identification of
the only known candidate for order parameter: the tetramerization
\cite{inami}.

Before proceeding, we would like to mention that \bavs{} has two
other phase transitions: a hexagonal-to-orthorhombic structural
transition at $T_{\rm S}=250{\rm K}$ \cite{massenet,Graf95}, and
the onset of antiferromagnetic long range order at $T_\chi=30{\rm
K}$ \cite{Graf95,Mihaly00,Nakamura00}. The role of the
orthorhombic splitting will be discussed elsewhere \cite{ivan},
while we neglect it in our discussions of the ARPES data. Our
results are taken at $T_\chi<T<T_{\rm S}$, on both sides of the
MIT.

The difficulties of interpreting the properties of \bavs{} are
partly due to the unclear character and occupancy of the relevant
$d$-states. In the ionic picture of a single V$^{4+}=3d^{1}$ site
surrounded by a distorted octahedron of S sites, the trigonal
component of the crystal field splits the $t_{2g}$ level into the
non-degenerate $a_{1g}$, and the doubly degenerate $e_{g}$ level.
If $e_{g}$ states are occupied, the interplay of orbital and spin
degrees of freedom is important.

Hopping processes broaden the levels into the corresponding
$a_{1g}$ and $e_g$ bands. Now the question is whether both bands
are partially occupied or it is sufficient to consider one band
only. If only the $a_{1g}$ band counts, the insulating phases of
\bavs{} must be describable in terms of atomic displacements and
spins. If only $e_g$ states are occupied, spin and orbital
ordering must be considered on an equal footing (a preliminary
discussion of this scenario was given in \cite{Mihaly00}). If both
$a_{1g}$ and $e_g$ electrons are present, complicated scenarios
can arise.

Choosing the orthorhombic $c$-axis as the $z$ axis, the $a_{1g}$
orbital has $z^2$ character, with strong overlap in the
$c$-direction. The $a_{1g}$ band is wide and almost
one-dimensional. If only the $a_{1g}$ band were filled, \bavs{}
could be an ordinary semiconductor or a weakly correlated metal,
without the possibility of localized moment formation. This is
obviously not the case. There must be also $e_g$ electrons. In
fact, all band structure calculations
\cite{Mattheiss95,Whangbo02,guo} agree that the Fermi level is
lying in the region of the crossing of $a_{1g}$-like and
$e_g$-like bands. However, the status of these predictions is
somewhat uncertain because neither of the calculations yields the
possibility of a non-magnetic insulating state. It is important to
have experimental results about the filling of the $e_g$ and
$a_{1g}$ bands, and the nature of the electronic structure in the
intermediate ($T_{\chi}<T<T_{\rm MI}$) insulating phase.

For many years, angle-resolved photoemission (ARPES) has been used to
experimentally probe the electronic band structure of different
materials. This powerful approach, however, requires high quality
single crystals of reasonable size. We were able to grow suitable
crystals for \bavs{} to experimentally determine the band
structure and therefore to provide a solid background for the
understanding of this compound. The most relevant result is the
indication that the MIT is associated with
interband nesting.

\begin{figure}
\includegraphics{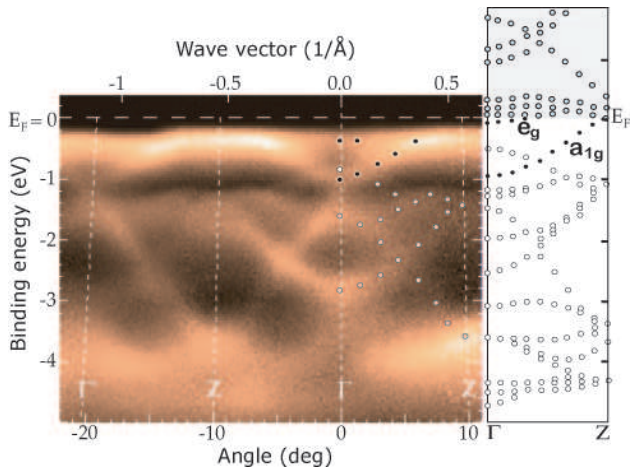}
\caption{\label{fig:arpesmap} (Left) ARPES intensity map taken in
the direction parallel to structural chains at 40K. Brighter
color signifies higher intensity. Bands arising from the V(3$d$)
level are followed with black circles, and the S(3$p$) bands with
the white ones. The spectra were normalized and a background was
subtracted to enhance the features. (Right) Corresponding theoretical band
dispersion from Ref. \cite{Mattheiss95}}
\end{figure}

We performed ARPES on fairly large ($\approx 0.25 \times 0.25
\times 3$ mm$^{3}$) single crystals of \bavs{} grown by the slow
cooling technique in melted Tellurium \cite{Kuriyaki00}. The data
were collected at the PGM beamline of the SRC, Stoughton (WI), USA, and with a Scienta-2002
analyzer. The spectra presented were measured with a total energy
and momentum resolution of $\Delta$E = 15meV and ${\Delta}k=$ 0.04
$\textrm{\AA}^{-1}$. Clean surfaces were exposed in UHV conditions
of the analyzer chamber (10$^{-11}$mbar range). The temperature of
the sample could be controlled in the range from 5K to 150K.

Due to the pronounced one-dimensional structure, samples were
fractured rather than cleaved and made momenta of electrons poorly defined in perpendicular-to-chain direction. The
measurements in the parallel-to-chain direction, however, were
successfully reproduced every time.

Fig. 1 shows the ARPES intensity map taken in the $\Gamma$--Z
direction parallel to the chains. For best resolution,
measurements were taken at T = 40K, and with a photon energy of
50eV, where we found all features close to the Fermi level $E_{F}$
distinctly resolved. A corresponding part of the \textit{ab
initio} calculation for the orthorhombic phase by Mattheiss
\cite{Mattheiss95} is displayed next to the map. The measured
intensity maps are essentially unchanged from 40K to 150K, except
for the temperature broadening of the features and the shift of
the leading edge close to the Fermi level as discussed later.

The theoretical $\textbf{\textit{k}}$-space periodicity in the
extended zone scheme is well reproduced in Fig.~1, indicating that
the measured surface gives a good probe for the bulk states. We
read off the period $2z$ where
$z=\overline{\Gamma\textrm{Z}}\approx\pi/c_{0}=0.56\textrm{\AA}^{-1}$,
$c_0= 5.61\textrm{\AA}$ being the $c$-axis lattice constant for
the two-atomic unit cell.

We find that both the positions and the widths of the bands at
energies above 1eV - originating from the S($3p$) orbitals - are
in agreement with the calculations \cite{Mattheiss95}.

Details of the electronic states just below the $E_{F}$ are shown
in the $-d^{2}I/dE^{2}$ intensity map of Fig. 2(c). The second
derivative clearly reveals two bands: a dispersive band with $\lesssim1$eV bandwidth,
which we identify
as a one-dimensional $a_{1g}$ band; and an apparently rather non-dispersive
band located at $\sim$0.4eV, which we associate with $e_g$
states. Selected raw measured energy distribution curves (EDCs)
presented in the Fig. 2(d) also demonstrate both features. This analysis only crudely determines the band structure close to the Fermi level. A different approach is needed, and discussed in the following paragraph, to resolve the structure in the closest proximity of the E$_{F}$, essential for the transport properties.

\begin{figure}
\includegraphics{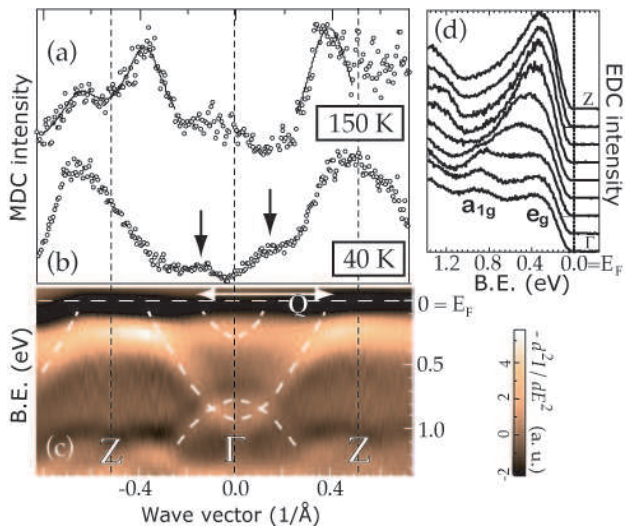}
\caption{\label{fig:crossing} (a) An MDC of 50meV wide integrated
region just below the Fermi energy ($E^{\ast}_{B}=50$meV) from an ARPES map at 150K. The fit is a double Lorentzian. (b) The equivalent MDC
plot ($E_{B}=E^{\ast}_{B}-80$meV) from a map at 40K. The arrows indicate the positions of $k_{F_{2}}$. (c)$-d^{2}I/dE^{2}$ plot of the ARPES map at 40K. The dashed lines serve as guides to the eye. (d) Selected raw EDC's in equidistant steps from $\Gamma$ to Z point.}
\end{figure}

A previous photoemission study reported the results of angle-integrated
measurements on polycrystalline samples \cite{Nakamura94}. Our
angle-resolved results on good single crystals confirm the
previous findings about the lack of a Fermi edge in a temperature
range above $T_{\rm MI}$. In our ARPES maps, peaks of all bands
fail to cross the Fermi level even in the metallic state. This was previously
attributed to Luttinger-liquid behavior.  However, the electronic
properties of \bavs{} are too far from being quasi-1D to permit such
an interpretation. On the other hand, we note that such
spectroscopic features are not unusual in correlated electron
systems (Ref \cite{bronze} and therein). Spectral weight, which is the only observable of an ARPES
experiment, is often renormalized by strong interactions which can
shift the spectral weight to higher binding energy masking the
real quasiparticle peak. Fermi level crossing is often still
observable with careful analysis. For that purpose we took a 
momentum distribution curve (MDC) from an ARPES
map taken at 150K with a 50meV wide window of integration around
the Fermi level. The obtained MDC shown in Fig. 2(a) reveals
peaks marking the crossings of the $a_{1g}$ band within the
first and the second Brillouin zone. The crossing determined from
Lorentzian peak fits, takes place at the wave vector $k_{F_{1}} =
(0.40 \pm 0.05) \textrm{ \AA}^{-1}$. The symmetry of the crossings from both sides of the BZ
positions the point $\textrm{Z} = (0.56 \pm 0.05) \textrm{ \AA}^{-1}$,
in agreement with the theoretical value.

The $e_g$-band is not so well resolved. However, the increase in
intensity not far from the $\Gamma$ point in Fig. 2(a) indicates
the possibility of a Fermi level crossing, creating a shallow
electron pocket. If we look at the equivalent MDC in the low
temperature map (Fig. 2(b)) that accounts for the spectral
changes due to the gap opening, the electron pocket shape of the
band is clearer. The band should cross $E_{F}$ at $k_{F_{2}} =
(0.15 \pm 0.05) \textrm{ \AA}^{-1}$. The electron pocket shape is
in agreement with the theoretical predictions.

The apparent bending of the $e_g$ band towards the $a_{1g}$ band
is a consequence of the resolution and back-folding of
\textit{shadow} bands -- the signature of lattice distortions
\cite{bronze}. Equally, the $a_{1g}$ band folds into shadow bands,
which is the reason behind the high intensity around the Z point
since the signals from both sides of the Brillouin zone boundary
overlap.

We cannot see the higher-lying part of the $a_{1g}$
band (see Fig.1 (right)) but its overall width must approach the theoretical value
$W(a_{1g})\sim 2$eV. On the other hand, we estimate $W({e_g}) <
0.4$eV. We conclude that the intriguing behavior of \bavs{}
results from the fact that two kinds of $d$-states: strongly
correlated narrow-band $e_g$-states, and weakly correlated
wide-band $a_{1g}$-states, coexist at the Fermi level. 

Clearly, the transport properties, including the observed low anisotropy, should originate from both bands that cross the E$_{F}$. Indeed, the
estimate of the ratio of averaged Fermi velocities \cite{Mattheiss95}, obtained supposing band structure similar to our findings, gives $\langle v_{\|}^{2}\rangle /\langle v_{\bot}^{2}\rangle \propto
{\sigma_c/\sigma_a}\sim 3.8$, in
close agreement with the experimental result \cite{Mihaly00}. New calculations based on our experimental data find the same ratio \cite{ivan}. In addition, our data show that the $a_{1g}$
band is hybridized with one of the sulphur $\pi$ bands, which
should play a role in reducing the macroscopic manifestation of
the quasi-1D crystal structure.

We ascribe the Curie-like susceptibility, and the bad metallic character of
the $T>T_{\rm MI}$ phase, to $e_g$ electrons, and to
$e_g$--$a_{1g}$ scattering. As to the symmetry breaking aspect of
the MIT, the orbital degrees of freedom of the $e_g$ electrons are
expected to give an electronic order parameter complementary to
the structurally defined tetramerization amplitude. Spin--orbital
models have the capacity to describe the development of a spin
gap, which is known to accompany the MIT of \bavs{} \cite{kezs01}.

Strong one-dimensional structural fluctuations are observed in a
wide temperature range above the MIT in diffraction experiments
\cite{fagot}. Such observations are common in quasi-1D charge
density wave (CDW)-bearing systems. However, a CDW is only one of the DW states \bavs{} can support, and the CDW is likely to be accompanied by the modulation of orbital character. Nevertheless, as in CDW systems, the 1D
fluctuations ought to have a relation to
${\textbf{\textit{k}}}$-space features. We examined our
experimentally determined band structure to check if we can
identify a Fermi surface instability which can introduce a gap
into the wide $a_{1g}$ band.

We find that the $a_{1g}$ band alone cannot satisfy the nesting
condition, thus we exclude intra-band nesting. We can also exclude
a 4$k_{F}$ instability, or the possibility of the $e_g$ band
supporting nesting in itself. On the other hand, our results are consistent with an
instability involving both bands, since our data are in agreement with a condition $k_{F_{1}} +
k_{F_{2}} = 0.5c^{\ast} = Q_{\rm CDW}$. The mechanism
of interband nesting leading to a CDW insulating state is known in
low-dimensional systems such as quasi-1D organic conductors and
K$_{0.3}$MoO$_{3}$ blue bronze. However, the nature of the
$T_{\chi}<T<T_{\rm MI}$ phase of \bavs{} is not yet clear, and the
precise mechanism of the instability involving both kinds of bands
remains to be identified.

\begin{figure}
\includegraphics{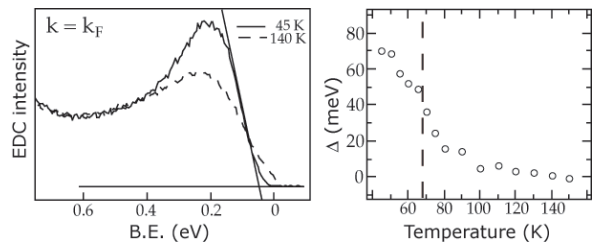}
\caption{\label{fig:transition} (Left) EDC's taken at 140 and 45K at the
E$_{F}$ crossing of the $a_{1g}$ band. The method of leading edge
position extraction is shown on the low-temperature EDC and (Right)
is its dependence on the temperature. The dashed line in (right) marks the MIT temperature.}
\end{figure}

We have monitored the spectral changes with temperature through
the MIT. The position of the peak in the
EDCs of our intensity maps (Fig.~3) stays at the same position but
the leading edge moves to higher binding energies. If we plot the
position of the leading edge versus temperature, we see a
monotonic shift of the leading edge - that most probably started
above 150K - and a noticeable non-linear increase in the shift
below 90K. This is consistent with the results of Ref.
\cite{Nakamura94} and indicative of the opening of a charge gap.
We do not see a clear transition associated with the T$_{\rm MI} =
69$K. This is not unusual, particularly when pre-transitional
fluctuations are present. The gap as seen in photoemission spectra
develops in the same range of temperature where resistivity
\cite{Mihaly00,forro00} and thermopower \cite{neven} measurements
identify a "precursor" to the insulating phase, and where X-Ray
diffraction detects large 1D fluctuations \cite{fagot}. The
saturation of the leading edge shift indicates a charge gap of
$\Delta _{\rm ch}=60-70{\rm meV}$. This is in good agreement with
the value obtained from transport measurements in the temperature
range where the Arrhenius law applies \cite{Graf95, Nakamura94,
Mihaly00}.


We point it out that the  gap opening can be seen for any choice
of $k$ vector, as should be expected if the complete Fermi surface
participates and if we take into account that the closeness of the
bands and surface-defined momentum resolution smear out signals
from both bands across the Brillouin zone.

The broad, pseudogapped spectral features are incompatible with a standard
quasiparticle picture. Recently, it was argued that in Peierls systems they are a
signature of polaronic carriers. The strong case for this scenario was found in
the studies on K$_{0.3}$MoO$_3$ blue bronze \cite{bronze}. Analogous considerations
may hold for \bavs{}, where the MIT has certain aspects of a spin-Peierls transition,
and the coupling of the  $e_g$ electrons to the local Jahn--Teller modes is important.
Interpretation of spectra in terms of Luttinger liquid or signal overlap from
multiple bands is unlikely, since similar consideration for blue bronze were
eliminated through studies on the related insulating, "one-band"
system K$_{0.33}$MoO$_{3}$ (red bronze) \cite{bronze}.

To conclude, we derived the structure of the bands close to the
Fermi level from ARPES measurements in a wide range of
temperatures on both sides of the MIT at
$69{\rm K}$. We found that the physics of {\bavs}  is
governed by the coexistence of weakly correlated wide-band
$a_{1g}$ electrons and strongly correlated narrow-band $e_g$
electrons near the Fermi level. Our results are consistent with
the $a_{1g}$--$e_g$ interband nesting condition and we propose it
as a plausible mechanism for the metal--insulator transition.

The authors are grateful for illuminating discussions to T.
Feh\'er and S. Bari\v{s}i\'{c}. Initial measurements were
performed at ISA, University
of Aarhus, Denmark, and we are grateful to P. Hofmann for this
opportunity. The SRC, UW - Madison, is supported by the NSF under
Award No. DMR-0084402. This work was supported by the Swiss
National Science Foundation through the MaNEP NCCR. P.F.
acknowledges support by the Hungarian National Grant T038162.

{\sl Note added}. The results of a recent LDA+DMFT calculation by F. Lechermann et al (cond-mat/0409463) about the role of correlations in determining the electron distribution over a$_{1g}$ states are in good overall agreement with our experimental findings.Ó

\end{document}